# On the Resistance of an Infinite Square Network of Identical Resistors
# (Theoretical and Experimental Comparison)

**J. H. Asad[†], A. Sakaji[††], R. S. Hijjawi[†††] and J. M. Khalifeh[†]**
[†]*Department of Physics, University of Jordan, Amman-11942, Jordan.*
[††] *Physics Department, Ajman University, UAE.*
[†††] *Physics Department, Mutah University, Jordan.*

**Abstract**

A review of the theoretical approach for calculating the resistance between two arbitrary lattice points in an infinite square lattice (perfect and perturbed cases) is carried out using the lattice Green's function. We show how to calculate the resistance between the origin and any other site using the lattice Green's function at the origin, $G_o(0,0)$, and its derivatives. Experimental results are obtained for a finite square network consisting of 30x30 identical resistors, and a comparison with those obtained theoretically is presented.

Key words: Lattice Green's Function, Resistors, Square lattice.

## 1. Introduction

It is an exiting question to find the resistance between two adjacent lattice points of an infinite square lattice where all the edges represent identical resistors R. This problem was studied well in many references[1-7], where for finding the resistance between two arbitrary grid points of an infinite square lattice they used a method based on the principle of superposition of current distributions[3-5]. One can find a full discussion for the electric circuit in van der Pol and Bremmer[1]. In the 1970's Montgomery[8] introduced a method for measuring the electrical resistivity of an isotropic material, where he prepared a rectangular prism with edges in principal crystal directions with electrodes on the corners. Based on Montgomery paper, Logan et. al[9] developed Montgomery's method, where they introduced potential series for computing the current flow in the rectangular block.

Recently Cserti[10] and Cserti et. al[11] studied the problem in which they used a alternative method based on the Lattice Green's Function (LGF) which enables us to calculate the resistance between any two arbitrary sites in a perfect and perturbed infinite square lattice.

The LGF has many applications in physics such as describing the interaction between the electrons which is mediated by the Phonons[12], studying the effect of impurities on the transport properties of metals[13], studying the transport in inhomogeneous conductors[14], studying the phase transition in classical two-dimensional lattice coulomb gases[15], and finally resistance calculation[10,11]. The LGF for the two-dimensional lattice has been studied well[16-19], and in these references the reader can find useful papers.

The LGF presented in this paper is related to the LGF of the Tight-Binding Hamiltonian (TBH)[13]. In the following we show how to calculate the resistance between the origin and any other site using $G_o(0,0)$ and its derivatives. The resistance between the origin and a lattice site $(l,m)$ in a constructed finite perfect square mesh (30x30 resistors) is measured. Also, the resistance between the origin and a lattice site $(l,m)$ in the same constructed mesh, when one of the resistors is broken (i.e. perturbed) is measured. Finally, a comparison is carried out between the measured resistances and those calculated by Cserti's method[10,11]. We believe that investigation of the resistance of square network of resistors should be interested in the field of arrays of Josephson junctions of high $T_C$ superconducting materials[20-22]. The investigation of electrophysical properties of such systems in the normal state before superconducting transition, and the content of this manuscript is helpful for electric circuit design and the method is instructive.

## 2. Theoretical Results

### 2.1 Perfect Square Lattice

In an infinite square lattice consisting of identical resistances R, the resistance between the origin and any lattice point $(l,m)$ can be calculated using[10]

$$R_o(l,m) = R[G_o(0,0) - G_o(l,m)] \tag{1}$$

where $G_o(0,0)$ is the LGF of the infinite square lattice at the origin, and $G_o(l,m)$ is the LGF at the site $(l,m)$.

First of all, the resistance between two adjacent points can easily be obtained as

$$R_o(1,0) = R[G_o(0,0) - G_o(1,0)]. \tag{2}$$

$G_o(1,0)$ can be expressed as (see Appendix A)

$$G_o(1,0) = \frac{1}{2}[tG_o(0,0) - 1] \quad ; \quad t = 2. \tag{3}$$

Where t is the energy.

and t=2 refers to the energy of the infinite square lattice at which the density of states (the imaginary parts of the LGF) is singular (Van Hove singularities) [23–25].

Thus, Eq. (2) becomes

$$R_o(1,0) = R[G_o(0,0) - G_o(0,0) + \frac{1}{2}] = \frac{R}{2}. \tag{4}$$

So, $R_o(1,0) = R_o(0,1) = \frac{R}{2}$. (due to the symmetry of the lattice).

The same result was obtained by Venzian[3], Atkinson et. al.[4], and Cserti[10].

To calculate the resistance between the origin and the second nearest neighbors (i.e. (1, 1)) then

$$R_o(1,1) = R[G_o(0,0) - G_o(1,1)]\,. \qquad (5)$$

$G_o(1,1)$ can be expressed in terms of $G_o(0,0)$ and $G'_o(0,0)$ as (see Appendix A)

$$G_o(1,1) = (\frac{t^2}{2} - 1)G_o(0,0) - \frac{t}{2}(4 - t^2)G'_o(0,0)\,. \qquad (6)$$

$$G_o(0,0) = \frac{2}{\pi t}K(\frac{2}{t}) \text{ and } G'_o(0,0) = \frac{-E(\frac{2}{t})}{\pi\, t(t-2)} - \frac{1}{\pi\, t^2}K(\frac{2}{t})\,. \qquad (7)$$

Where $K(\frac{2}{t})$ and $E(\frac{2}{t})$ are the elliptic integrals of the first kind and second kind respectively.

Substituting the last two expressions into Eq. (4), one obtains

$$R_o(1,1) = \frac{2R}{\pi}\,. \qquad (8)$$

Again our result is the same as Cserti[10] and Venezain[3].

Finally, to find the resistance between the origin and any lattice site $(l,m)$ one can use the above method, or we may use the recurrence formulae presented by Cserti [10] (i.e. Eq. 32).

So, using Eq. (32) in Cserti[10] and the known values of $R_o(0,0) = 0$, $R_o(1,0) = \frac{R}{2}$ and $R_o(1,1) = \frac{2R}{\pi}$ we calculate exactly the resistance for arbitrary sites. The same result was obtained by Atkinson et. al[4], and below are some calculated values:
$\frac{R_o(2,0)}{R} = 0.7267$, $\frac{R_o(3,0)}{R} = 0.8606$, and $\frac{R_o(4,0)}{R} = 0.9539$.
For large values of $l$ or/and $m$ the resistance between the origin and the site $(l,m)$ is given as[10]

$$R_o(l,m) = \frac{R}{\pi}(Ln\sqrt{l^2 + m^2} + \gamma + \frac{Ln8}{2}) \qquad (9)$$

where $\gamma = 0.5772$ is the Euler-Mascheroni constant[26]. Venezain obtained the same result[3].
Finally, as *l* or *m* goes to infinity then the resistance in a perfect infinite square lattice divergence.

## 2.2 Perturbed Square Lattice (a bond is broken)

The resistance between the sites $i$ and $j$ of the perturbed infinite square lattice where the bond between the sites $i_o$ and $j_o$ is broken can be calculated using[11]

$$R(i,j) = R_o(i,j) + \frac{[R_o(i,j_o) + R_o(j,i_o) - R_o(i,i_o) - R_o(j,j_o)]^2}{4[R - R_o(i_o,j_o)]}. \qquad (10)$$

where $i = (i_x, i_y)$, $J = (j_x, j_y)$, $i_o = (i_{ox}, i_{oy})$ and $j_o = (j_{ox}, j_{oy})$.
The resistance between the ends of the removed bond (i.e. $R(i_o, j_o)$) is equal to $R$ [11].

To calculate the resistance in the perturbed network, one has to specify clearly the ends of the removed bond. For example, when the broken bond is taken to be between the sites $i_o = (0,0)$ and $j_o = (1,0)$ we found using Eq. (10) that:
$\frac{R(1,0)}{R} = 1.0000$, $\frac{R(2,0)}{R} = 0.9908$, $\frac{R(3,0)}{R} = 1.0614$, and $\frac{R(4,0)}{R} = 1.1300$.
Now, if the broken bond is shifted and taken to be between the sites $i_o = (1,0)$ and $j_o = (2,0)$ we found again using Eq. (10) that:
$\frac{R(1,0)}{R} = 0.5373$, $\frac{R(2,0)}{R} = 0.9908$, $\frac{R(3,0)}{R} = 0.9634$, and $\frac{R(4,0)}{R} = 1.0189$.
For large separation between the two sites, then

$$R(i,j) = R_o(i,j) + \frac{\left[\frac{R}{\pi} Ln \sqrt{\frac{i^2 j^2 + i^2 i_o^2 + j_o^2 j^2 + j_o^2 i_o^2}{i^2 j^2 + i^2 j_o^2 + i_o^2 j^2 + i_o^2 j_o^2}}\right]^2}{4[R - R_o(i,j)]}. \qquad (11)$$

Now, as $i$ or/and $j$ goes to infinity then

$$R(i,j) \to R_o(i,j). \qquad (12)$$

that is, the perturbed resistance between arbitrary sites goes to the perfect resistance as the separation between the two sites goes to infinity.

.

## 3. Experimental Results

To study the resistance of a finite square lattice experimentally we constructed a finite square network of identical ($30\times 30$) carbon resistors, each have a value of ($1\ K\Omega$) and a tolerance of (1%).

### 3.1 Perfect Case

Using the constructed network, the resistance between the origin and the site *(l,m)* is measured (using two-point probe). Below are some measured values:

$\frac{R_o(1,0)}{R}=0.4997$, $\frac{R_o(2,0)}{R}=0.7283$, $\frac{R_o(3,0)}{R}=0.8642$, and $\frac{R_o(4,0)}{R}=0.9616$.

The above measured values are very close to those calculated in section 2.1.

### 3.2 Perturbed Case

In this section the bond between the sites $i_o=(0,0)$ and $j_o=(1,0)$ is removed and the resistance between the sites $i=(0,0)$ and $j=(j_x,j_y)$ is measured using the same network. Below are some measured values:

$\frac{R(1,0)}{R}=1.0020$, $\frac{R(2,0)}{R}=0.9939$, $\frac{R(3,0)}{R}=1.0670$, and $\frac{R(4,0)}{R}=1.1410$.

Now, the broken bond is shifted to be between the sites $i_o=(1,0)$ and $j_o=(2,0)$. Again we measure the resistance between the sites $i=(0,0)$ and $j=(j_x,j_y)$, and below are some measured values:

$\frac{R(1,0)}{R}=0.5372$, $\frac{R(2,0)}{R}=0.9939$, $\frac{R(3,0)}{R}=0.9689$, and $\frac{R(4,0)}{R}=1.0290$.

As shown, one can see that there is an excellent agreement between the measured and the calculated values.

## 4. Results and Discussion

From the Figures shown the resistance in an infinite square lattice is symmetric under the transformation $(l,m)\rightarrow(-l,-m)$ due to the inversion symmetry of the lattice. However, the resistance in the perturbed infinite square lattice is not symmetric due to the broken bond, except along the [01] direction since there is no broken bond along this direction.

Also, one can see that the resistance in the perturbed infinite square lattice is always larger than that in a perfect lattice and this is due to the positive second term in Eq. (10). But as the separation between the sites increases the perturbed resistance goes to that of a perfect lattice.

The constructed mesh gives accurately the bulk resistance shown in Figs. 3.1-3.6, and this means that a crystal consisting of (30x30) atoms enables one to study the bulk properties of the crystal in a good way. But, as we approach the edge then the measured resistance exceeds the calculated one and this is due to the edge effect. Also, one can see from the figures that the measured resistance is symmetric in the perfect mesh, which is expected.

Fig. 3.4 and Fig. 3.6 show that the measured resistance along the [01] direction is nearly symmetric within experimental error, which is expected due to the fact that there is no broken bond along this direction, and this is in agreement with the theoretical result.

Finally, our values are in good agreement with the bulk values calculated by Cserti's method[10,11] . Derivation of the resistance of a finite square lattice is under investigation in order to compare with the realistic experimental results.

# Figure Captions

**Fig. 3.1 The resistance between** $i = (0,0)$ **and** $j = (j_x, 0)$ **of the perfect square lattice as a function of** $j_x$ **; calculated (squares) and measured (circles) along the [10] direction.**

**Fig. 3.2 The resistance between** $i = (0,0)$ **and** $j = (j_x, j_y)$ **of the perfect square lattice as a function of** $j_x$ **and** $j_y$ **; calculated (squares) and measured (circles) along the [11] direction.**

**Fig. 3.3 The resistance between** $i = (0,0)$ **and** $j = (j_x, 0)$ **of the perturbed square lattice as a function of** $j_x$ **; calculated (squares) and measured (circles) along the [10] direction. The ends of the removed bond are** $i_o = (0,0)$ **and** $j_o = (1,0)$ **.**

**Fig. 3.4 The resistance between** $i = (0,0)$ **and** $j = (0, j_y)$ **of the perturbed square lattice as a function of** $j_y$ **; calculated (squares) and measured (circles) along the [01] direction. The ends of the removed bond are** $i_o = (0,0)$ **and** $j_o = (1,0)$ **.**

**Fig. 3.5 The resistance between** $i = (0,0)$ **and** $j = (j_x, 0)$ **of the perturbed square lattice as a function of** $j_x$ **; calculated (squares) and measured (circles) along the [10] direction. The ends of the removed bond are** $i_o = (1,0)$ **and** $j_o = (2,0)$ **.**

**Fig. 3.6 The resistance between** $i = (0,0)$ **and** $j = (0, j_y)$ **of the perturbed square lattice as a function of** $j_y$ **; calculated (squares) and measured (circles) along the [01] direction. The ends of the removed bond are** $i_o = (1,0)$ **and** $j_o = (2,0)$ **.**

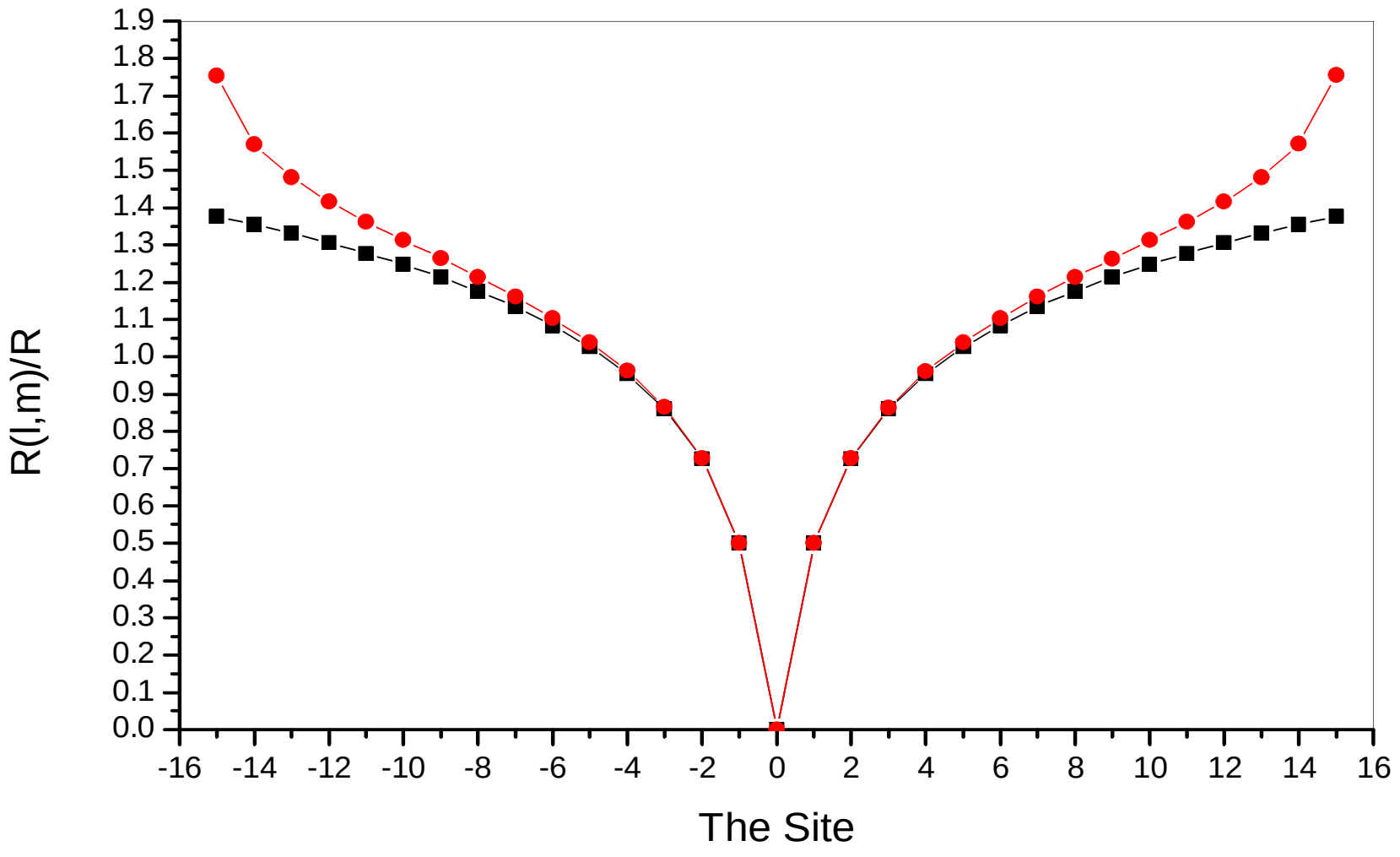

**Fig. 3.1**

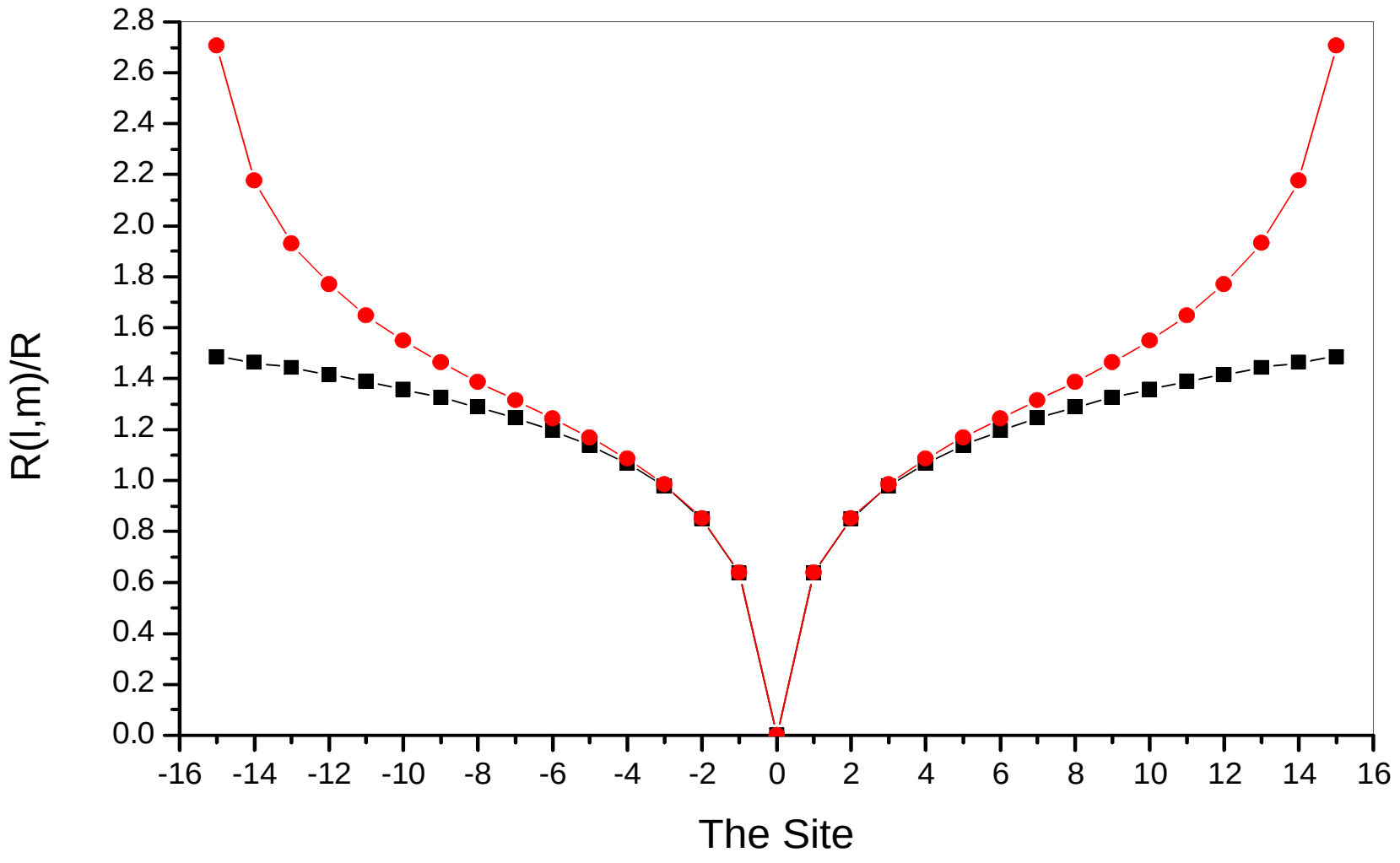

**Fig. 3.2**

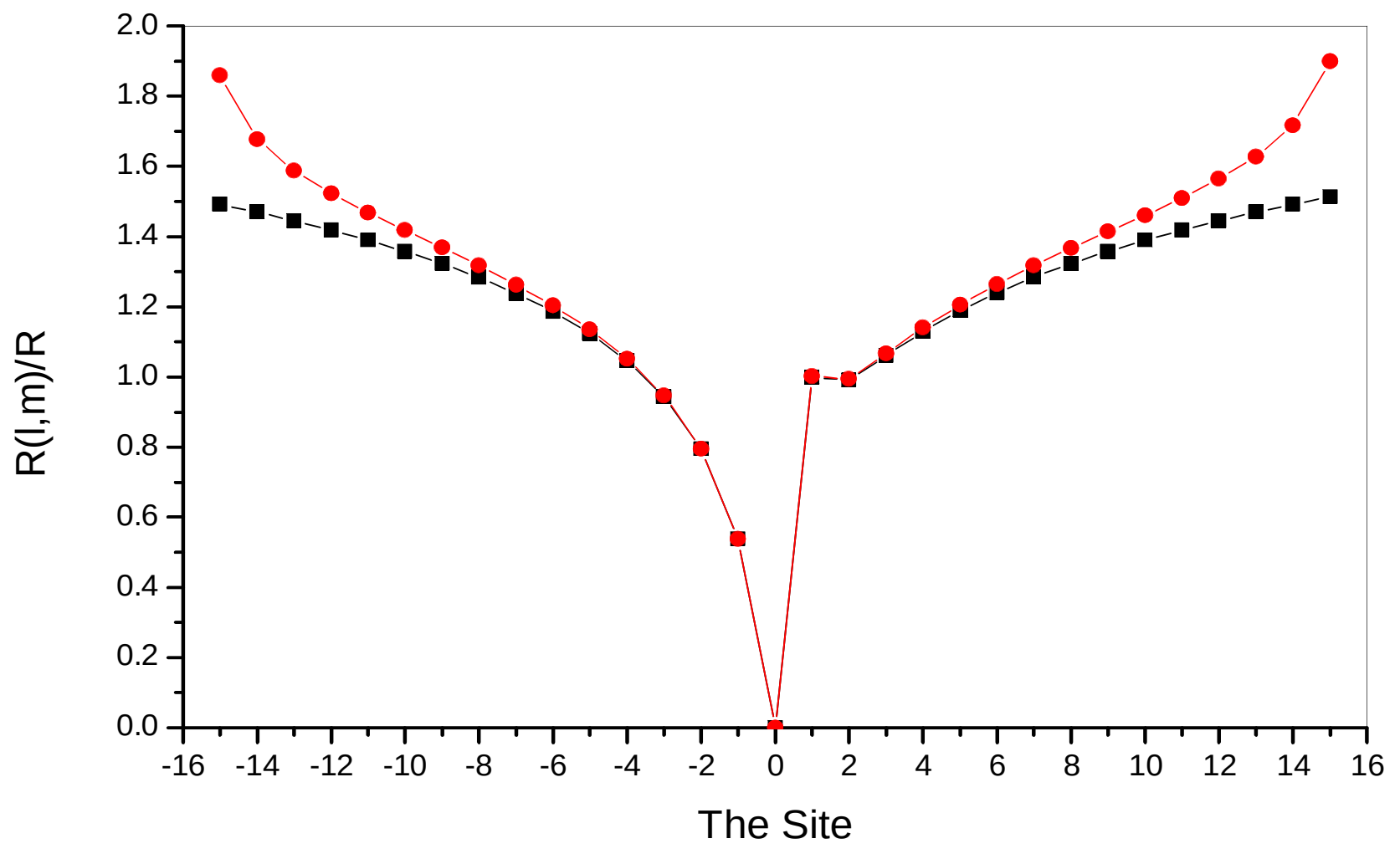
R(l,m)/R
The Site
2.0
1.8
1.6
1.4
1.2
1.0
0.8
0.6
0.4
0.2
0.0
-16 -14 -12 -10 -8 -6 -4 -2 0 2 4 6 8 10 12 14 16

**Fig. 3.3**

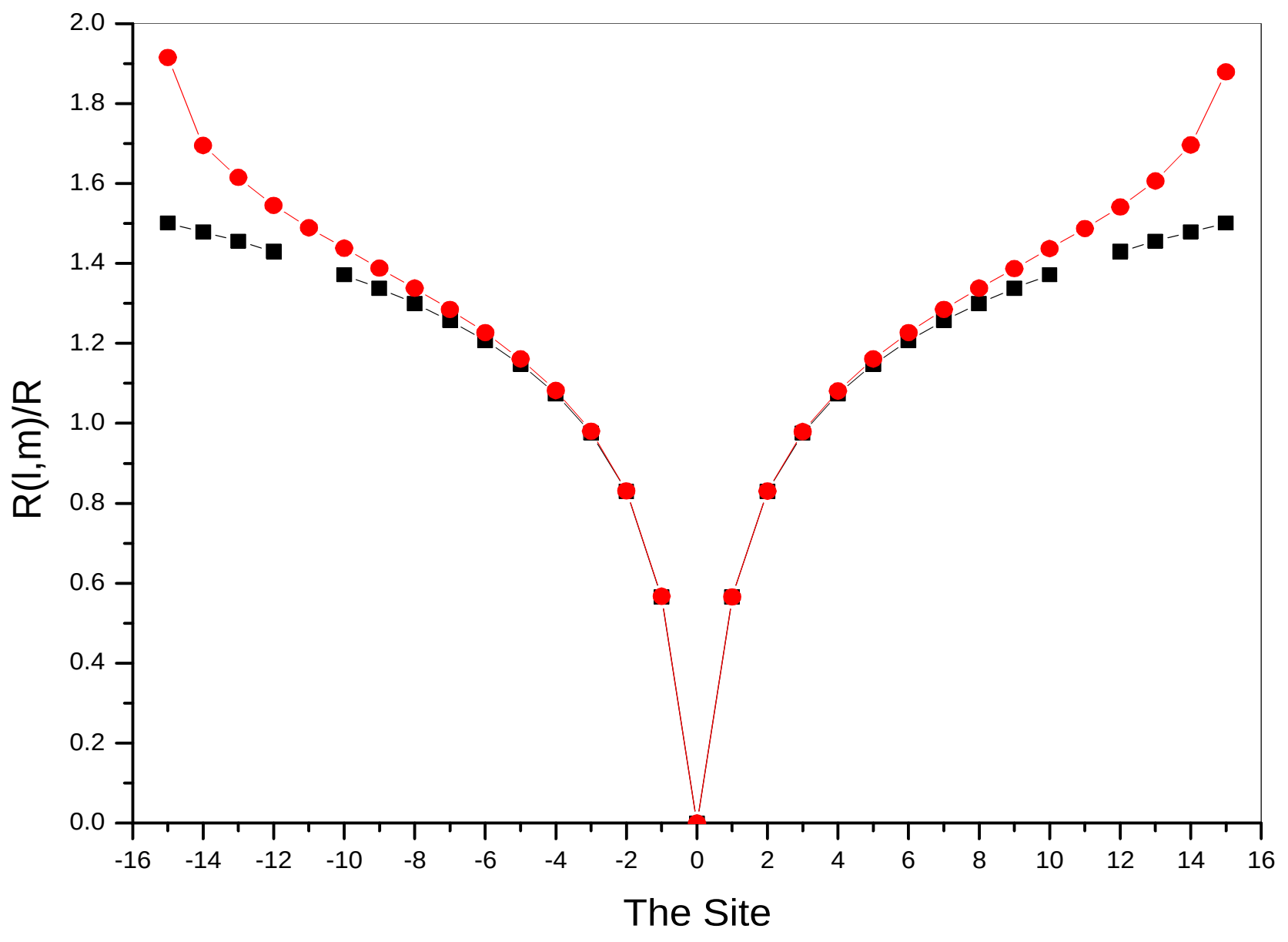
R(l,m)/R
The Site
2.0
1.8
1.6
1.4
1.2
1.0
0.8
0.6
0.4
0.2
0.0
-16 -14 -12 -10 -8 -6 -4 -2 0 2 4 6 8 10 12 14 16

**Fig. 3.4**

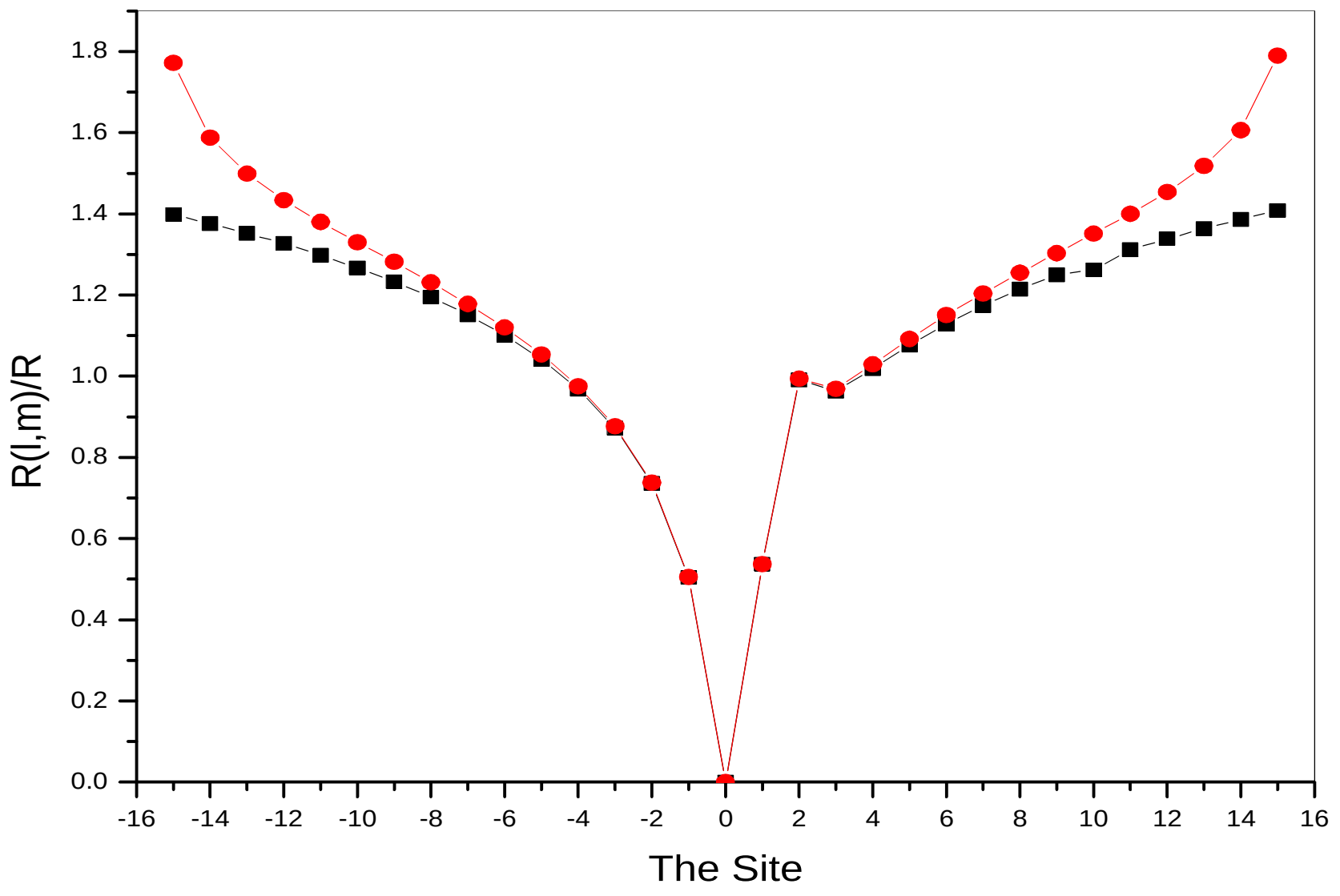
R(l,m)/R
The Site
1.8
1.6
1.4
1.2
1.0
0.8
0.6
0.4
0.2
0.0
-16 -14 -12 -10 -8 -6 -4 -2 0 2 4 6 8 10 12 14 16

**Fig. 3.5**

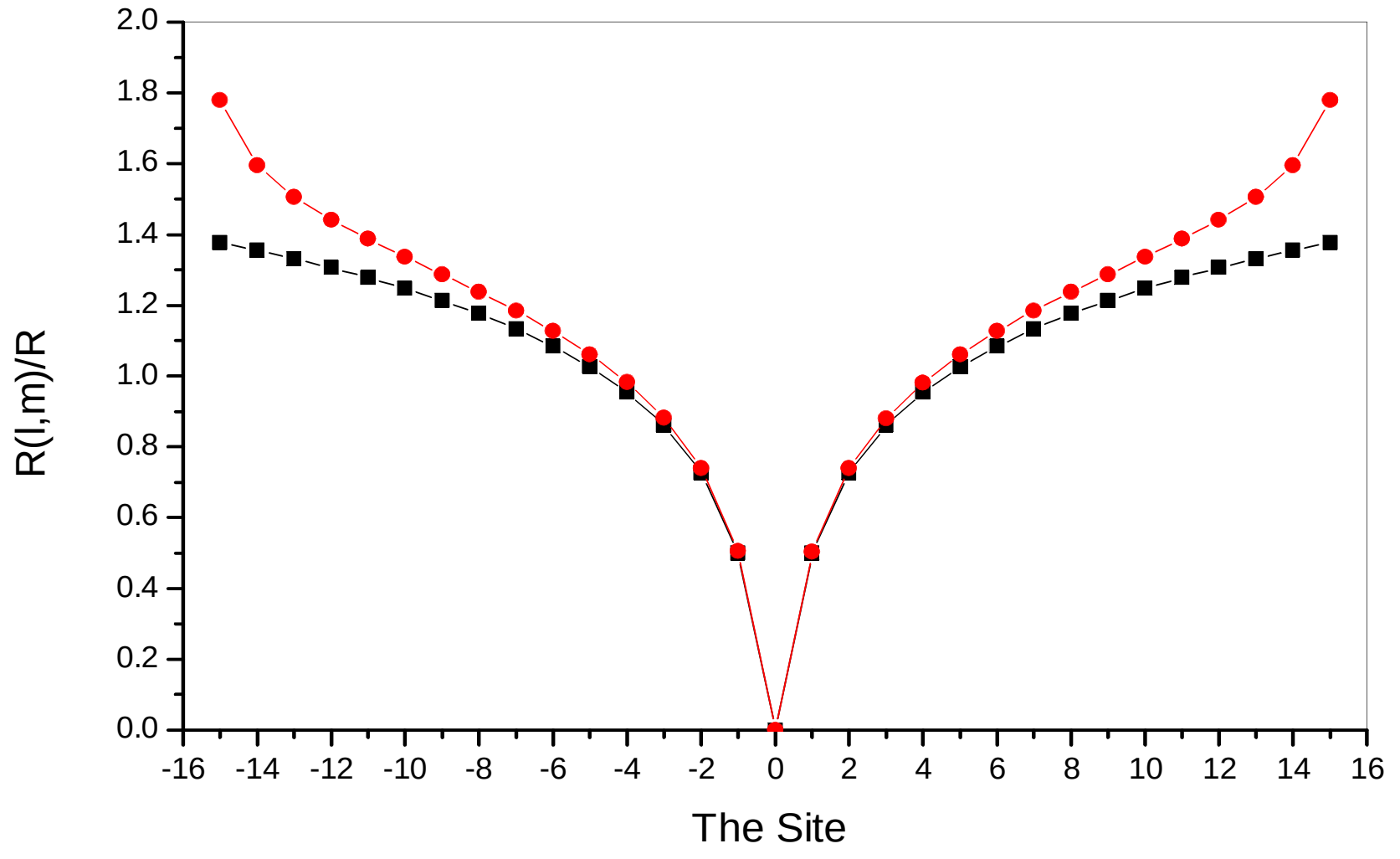
R(l,m)/R
The Site
2.0
1.8
1.6
1.4
1.2
1.0
0.8
0.6
0.4
0.2
0.0
-16 -14 -12 -10 -8 -6 -4 -2 0 2 4 6 8 10 12 14 16

**Fig. 3.6**

## APPENDIX A

The LGF for two-dimensional lattice is defined by[11]

$$G(m,n,t) = \frac{1}{\pi^2}\int_0^{\pi}\int_0^{\pi} \frac{\text{Cosmx Cosny}}{t - (\text{Cosx} + \text{Cosy})}\, dx\, dy, \qquad \text{(A1)}$$

Where (m, n) are integers and t is a parameter.

By executing a partial integration with respect to x in Eq. (A1), we obtained the following recurrence relation[20]:

$$G'(m+1,n) - G'(m-1,n) = 2mG(m,n), \qquad \text{(A2)}$$

where $G^{/}(m,n)$ expresses the first derivative of G(m,n) with respect to t. Taking derivatives of Eq. (A2) with respect to t, we obtained recurrence relations involving higher derivatives of the GF.

Putting (m,n)=(1,0), (1,1), and (2,0) in Eq. (A2), respectively we obtained the following relations:

$$G'(2,0) - G'(0,0) = 2G(1,0), \qquad \text{(A3)}$$

$$G'(2,1) - G'(1,0) = 2G(1,1), \qquad \text{(A4)}$$

$$G'(3,0) - G'(1,0) = 4G(2,0). \qquad \text{(A5)}$$

For m=0 we obtain[10,21-22]

$$2tG(0,n) - 2\delta_{0n} - 2G(1,n) - G(0,n+1) - G(0,n-1) = 0 \qquad \text{(A6)}$$

Insert n=0 in Eq. (A6) we find the well-known relation

$$G(1,0) = \frac{1}{2}[tG(0,0) - 1], \qquad \text{(A7)}$$

for m≠0 we have

$$G(m+1,n)-2tG(m,n)+G(m-1,n)+G(m,n+1)+G(m,n-1)=0, \tag{A8}$$

Substituting (m,n)=(1,0), (1,1), and (2,0) in Eq. (A8), respectively we obtained the following relations:

$$G(1,1)=tG(1,0)-\frac{1}{2}G(0,0)-\frac{1}{2}G(2,0), \tag{A9}$$

$$G(2,1)=(t^2-1)G(1,0)-\frac{t}{2}G(0,0)-\frac{t}{2}G(2,0), \tag{A10}$$

$$G(3,0)=(\frac{3}{2}t-t^3)G(0,0)+3tG(2,0)-(\frac{1-2t^2}{2}), \tag{A11}$$

Now, by taking the derivative of both sides of Eq. (A11) with respect to t, and using Eqs. (A3), (A4) and (A5), we obtained the following expressions:

$$G(2,0)=(4t-t^3)G'(0,0)+G(0,0)-t, \tag{A12}$$

$$G(1,1)=(\frac{t^2}{2}-1)G(0,0)-\frac{t}{2}(4-t^2)G'(0,0), \tag{A13}$$

$$G(2,1)=\frac{t}{2}(t^3-3)G(0,0)-\frac{t^2}{2}(4-t^2)G'(0,0)-\frac{1}{2}, \tag{A14}$$

$$G(3,0)=\frac{t}{2}(9-2t^2)G(0,0)+3t^2(4-t^2)G'(0,0)-(\frac{1+4t^2}{2}), \tag{A15}$$

Again, taking the derivative of both side of Eq. (A12) with respect to t, and using Eqs. (A3) and (A7), we obtained the following differential equation for G(0,0):

$$t(4-t^2)G''(0,0)+(4-3t^2)G'(0,0)-tG(0,0)=0. \tag{A16}$$

Where $G^{//}(0,0)$ is the second derivative of G(0,0).

By using the following transformations G(0,0) = Y(x)/t and $x=4/t^2$ we obtain the following differential equation [23-25] :

$$x(1-x)\frac{d^2Y(x)}{dx^2}+(1-2x)\frac{dY(x)}{dx}-\frac{1}{4}Y(x)=0, \qquad \text{(A17)}$$

This is called the hypergeometric differential equation (Gauss's differential equation). So, the solution is [25]

$Y(x) = {}_1F_2(1/2,1/2;1;x) = (2/\pi)\, K(2/t)$

Then,

$$G(0,0,t)=\frac{2}{\pi t}K\left(\frac{2}{t}\right) \qquad \text{(A18)}$$

By using Eq. (A18) we can express $G^{/}(0,0)$ and $G^{//}(0,0)$ in terms of the complete elliptic integrals of the first and second kind.

$$G'(0,0,t)=\frac{2}{\pi}\frac{E\left(\frac{2}{t}\right)}{4-t^2}, \qquad \text{(A19)}$$

$$G''(0,0,t)=\frac{2}{\pi t(t^2-4)}\left[E\left(\frac{2}{t}\right)[3t^2-4]-K\left(\frac{2}{t}\right)\right]. \qquad \text{(A20)}$$

K(2/t) and E(2/t) are the complete elliptic integrals of the first and second kind, respectively. So that, the two-dimensional LGF at an arbitrary site is obtained in closed form, which contains a sum of the complete elliptic integrals of the first and second kind.